\documentclass[12pt]{iopart}
\usepackage{times,dsfont,amssymb,amsmath,amsthm,amsfonts,amsbsy,mathrsfs,bm,bbm,graphicx,graphics,epsfig,multirow,mathtools}
\usepackage{iopams}
\usepackage{pdfsync}

\newcommand{\bra}[1]{\left\langle #1\right|}
\newcommand{\ket}[1]{\left| #1\right\rangle}

\newcommand{\ketbra}[3][]{\left|#2\right\rangle_{#1}\!\left\langle#3\right|}

\newcommand{\eeqref}[1]{Eq.~(\ref{#1})}

\newcommand{\ignore}[1]{}

\begin{document}

\title{Quantum process tomography with coherent states}
\author{Saleh Rahimi-Keshari$^{1}$, Artur Scherer$^{1}$, Ady Mann$^{1,2}$,  Ali T. Rezakhani$^{3,4}$,
  A.~I.~Lvovsky$^{1}$, and Barry C.\ Sanders$^1$}

\address{$^1$Institute for Quantum Information Science and Department of Physics and
Astronomy, University of Calgary, Alberta, Canada T2N 1N4}
\address{$^2$Physics Department, Technion, Haifa 32000, Israel}
\address{$^3$Department of Chemistry, Center for Quantum Information Science and Technology, University
  of Southern California, Los Angeles, California 90089, USA}
\address{$^4$Department of Physics, Sharif University of Technology, P. O. Box 11155-9161,
Tehran, Iran}
\ead{lvov@ucalgary.ca}

\begin{abstract}
We develop an enhanced technique for characterizing quantum optical
processes based on probing unknown quantum processes only with coherent
states. Our method substantially improves the original proposal
[M. Lobino et al., Science {\bf 322}, 563 (2008)], which
uses a filtered Glauber-Sudarshan decomposition to
determine the effect of the process on an arbitrary state.
We introduce a new relation between the action of
a general quantum process on coherent state inputs
and its action on an arbitrary quantum state.
This relation eliminates the need to invoke the Glauber-Sudarshan
representation for states; hence it dramatically simplifies
the task of process identification and removes a potential source
of error. The new relation also enables straightforward extensions of the method to multi-mode and non-trace-preserving processes. We illustrate our formalism
with several examples, in which we derive analytic
representations of several fundamental quantum optical processes
in the Fock basis. In particular, we introduce photon-number cutoff
as a reasonable physical resource limitation and address resource vs accuracy trade-off in practical applications. We show that the accuracy of process
estimation scales inversely with the square root of photon-number
cutoff.

\end{abstract}
\pacs{03.65.Wj, 42.50.-p, 03.67.-a.}
\maketitle
\section{Introduction}

Assembling a complex quantum optical information processor requires precise
knowledge of the properties of each of its components, i.e., the ability to
predict the effect of the components on an arbitrary input state. This gives
rise to a quantum version of the famous \lq\lq black box problem'', which is
addressed by means of {\it quantum process tomography} (QPT) \cite{Zoller,D'Ariano, Lidar}. In
QPT, a set of probe states is sent into the black box
(here an unknown completely-positive, linear quantum process $\mathcal{E}$
over the set of bounded operators $\mathcal{B}(\mathcal{H})$ on a Hilbert
space $\mathcal{H}$)
and the output states are measured. From the effect of
the process on the probe states it is possible to predict its effect on
any other state within the same Hilbert space.

QPT exploits linearity of quantum process over its density
operators. If the effect of the process
$\mathcal{E}(\rho_i)$ is known
for a set of density operators $\{\rho_i\}$, its effect on any linear
combination $\rho=\sum \beta_i\rho_i$ equals $
\mathcal{E}(\rho)=\sum \beta_i \mathcal{E}(\rho_i)$. Therefore, if
$\{\rho_i\}$ forms a spanning set within the space
$\mathcal{L}(\mathcal{H})$ of  linear operators over a particular Hilbert
space $\mathcal{H}$, knowledge of $\{ \mathcal{E}(\rho_i)\}$
is sufficient to extract complete information about the quantum process.

However, practical implementations of QPT become demanding especially for
systems with large Hilbert spaces.
For $\dim(\mathcal{H})=d$, $\dim\bigl(\mathcal{L}(\mathcal{H})\bigr)=d^2$,
which implies that at least $d^2$ unknown operators $\{\mathcal{E}(\rho_i)\}$, each with $d^2$ unknown parameters, must be estimated. This procedure requires preparation of at least $\{\rho_i\}_{i=1}^{d^2}$ states, subjecting each to the unknown process $\mathcal{E}$, and determining each element of $\{\mathcal{E}(\rho_i)\}_{i=1}^{d^2}$ through measurement
(each with $d^2$ unknown elements), thereby inferring an overall number
of $d^4$ parameters. Furthermore, in order to build up sufficient statistics
for reliable estimates of the output states, each measurement should be performed
many times on multiple copies of the inputs. Thus,
a large number of realizations and measurements are
required for complete tomography of $\mathcal{E}$.

An additional complication, especially for QPT of quantum {\it optical} processes,
is associated with preparation of the probe states. Typical
optical QPT implementations deal with systems consisting of one or more dual-rail qubits
\cite{O'Brien,Leung,Steinberg2},
which implies that the probe states are highly nonclassical,
hence difficult to generate.

These difficulties have been partially alleviated in the recently proposed
scheme of \lq\lq coherent-state quantum process tomography" (csQPT)
\cite{LobinoQPT}. This scheme is based on the observation that
the density operator $\rho$ of a generic quantum state of every
electromagnetic mode can be expressed in the {\em Glauber-Sudarshan representation}~\cite{Glauber,Sudarshan},
\begin{equation}
\rho=\int_{\mathds{C}} \mathrm{d}^2\alpha~ P_\rho(\alpha)|\alpha\rangle\langle\alpha|,
\label{eq.rho}
\end{equation}
where $P_\rho(\alpha)$ is a quasi-probability distribution referred to as the
quantum state's \lq\lq $P$-function''  and integrated over
the entire complex plane \cite{footnote1}.
Linearity hence implies that measuring
\begin{equation}
|\alpha\rangle\langle
\alpha|\mapsto \varrho_{\mathcal{E}}(\alpha)\equiv \mathcal{E}(\ketbra\alpha\alpha),
\end{equation}
i.e., determining the effect of the unknown process on {\em all} coherent
states enables a prediction of its effect upon any generic state $\rho$
according to
\begin{equation}\label{eq.edec}
   \mathcal E( \rho)=\int_{\mathds{C}}\mathrm{d}^2\alpha~
    P_\rho(\alpha)\varrho_{\mbox{\tiny $\mathcal{E}$}}(\alpha).
\end{equation}

The implementation of csQPT is advantageous because (i) coherent states are
readily available from lasers,
(ii) coherent states of different amplitudes and phases can be produced
without changing the layout of the experimental apparatus,
and (iii) output-state characterization can be performed using optical
homodyne tomography \cite{LvovskyRaymer}, which obviates the need for
postselection and provides full information about the process in question. Moreover, csQPT has been tested experimentally on simple
single-mode processes, such as the identity, attenuation, and phase
shift operations~\cite{LobinoQPT}. Furthermore, csQPT
has been used to characterize quantum memory for light
based on electromagnetically-induced transparency \cite{LobinoQPT2}.

An apparent obstacle to csQPT, however, is that the $P$
function for many nonclassical optical states exists only
in terms of a highly singular generalized function \cite{Cahill1,Cahill3}.
A remedy therefor is provided by Klauder's theorem~\cite{Klauder},
which states that any
trace-class operator
$\rho$
can be approximated, to arbitrary accuracy, by a bounded operator ${\rho_L}\in\mathcal{B}(\mathcal{H})$
whose Glauber-Sudarshan function $P_L$ is in the Schwartz class \cite{Gelfand1}, so integration \eqref{eq.edec} can be performed. The Klauder approximation
can be constructed by low-pass filtering of the $P$ function,
i.e., by multiplying its Fourier transform with
 an appropriate
regularizing function equal to $1$ over a square domain
of size $L\times L$ and rapidly dropping to zero outside this domain. Ref.~\cite{LobinoQPT} employs this method to implement csQPT.

Practical implementation of Klauder's procedure is however complicated, because it requires finding the characteristic function of the input state and subsequently its regularized $P$ function. This function features high-frequency, high-amplitude oscillations that limit the precision in calculating the output state \eqref{eq.edec}. Furthermore, Klauder's approximation is ambiguous in the choice of  the particular
filtering function as well as the cutoff parameter
$L$~\cite{LobinoQPT}.

Here we improve csQPT to overcome the above problems.
Specifically, we develop a new method for csQPT that eliminates the explicit use of
the Glauber-Sudarshan representation and thus removes the inherent ambiguity associated with employing Klauder's approximation for csQPT. In Sec.~\ref{SubSec:csQPT-Formalism}, we obtain an expression for the process tensor in the Fock (photon number) basis that can be directly calculated from the experimental data. Using this tensor, the process output for an arbitrary input can be calculated by simple matrix multiplication rather than requiring integration and high-frequency cut-offs. In this way, transformations between the Fock and Glauber-Sudarshan representations, which were necessary in Ref.~\cite{LobinoQPT}, can be sidestepped.
Using our new approach,
we easily extend csQPT from its restrictive single-mode applicability
to multi-mode processes and
even to non-trace-preserving conditional processes.
These extensions are particularly relevant
for quantum information processing circuits, whose basic components are
inherently multi-mode and conditional~\cite{KLM}.

Process tomography is successful if, for every input state, the estimate
for the process output closely approximates the actual process output
state, and the worst-case error of this estimate, given by a distance
between the actual and estimated process outputs, is less than a given
tolerance. For states over infinite-dimensional Hilbert spaces, this
concept of error is however not meaningful because the finiteness of
sampling implies that the process is necessarily under-sampled, hence
cannot be determined with bounded error. Instead we could consider the
process estimation restricted to a finite-dimensional \emph{subset} of
$\mathcal B(\mathcal H)$. This version of process tomography can always
be successful with a sufficiently large amount of sampling.

Of particular practical interest is the subspace $\mathcal B(\tilde{\mathcal H})$
defined by an energy cut-off, i.e., estimating
the process without accessing any information about its high-energy
behavior. This restriction is naturally consistent with our choice to
work in the Fock basis, because then the resulting process tensor is of
finite size and with many practical settings (e.g. quantum-information
processing with photonic qubits). In Sec.~\ref{SubSec:ErrorAnalysis}, we provide process error
estimates for several input state subsets that extend beyond $\mathcal
B(\tilde{\mathcal H})$.

Many interesting processes are phase symmetric; that is, an optical phase shift of the input state results in the same phase shift of the output. This property dramatically simplifies the experiment because one needs to collect data only for coherent states whose amplitudes lie on the real axis rather than the entire complex plane. This prompts us to discuss, in Sec.~\ref{Sec:ExperimentalApplication}, how to obtain the process tensor
for phase-symmetric processes, which we test on the experimental data from Ref.~\cite{LobinoQPT2}. Next, in Sec.~\ref{Sec:Examples}, we illustrate our method by analytically deriving the superoperators for certain fundamental quantum optical processes using the Fock basis. The paper is concluded in Sec.~\ref{conc} and is supplemented with two appendices.

\section{Coherent state quantum process tomography }
\label{Sec:csQPT}

\subsection{Formalism: determining the quantum process matrix}
\label{SubSec:csQPT-Formalism}

We study general quantum optical processes $\mathcal{E}$ acting on quantum
states of light and begin with the simplest case for which only a single
electromagnetic field mode is involved.
An arbitrary quantum state
$\rho$
can be expressed in the Fock
representation as
\begin{equation}
 \rho=\sum_{m,n=0}^{\infty}\rho_{mn}\ket{m}\bra{n}.
\label{Fock-Basis-ExpansionInputState}
\end{equation}
Subjecting this state to an unknown process $\mathcal{E}$, and imposing linearity, yields
\begin{equation}
\mathcal{E}(\rho)=\sum_{j,k,m,n=0}^{\infty} \rho_{mn}~\mathcal{E}^{mn}_{jk}|j\rangle \langle k|,
\label{Fock-Basis-ExpansionOutputState}
\end{equation}
where
\begin{equation}
\mathcal{E}^{mn}_{jk}:=\langle j\lvert\mathcal{E}(\lvert m\rangle\langle
n\rvert)\rvert k\rangle
\label{superoperator}
\end{equation}
is a rank--$\mathit{4}$ tensor, hereafter referred to as the \lq\lq process tensor" (superoperator).
Thus, by expressing input and output states in the Fock basis,
a quantum process can be  uniquely represented and characterized
by its rank-$\mathit{4}$ tensor, which relates the
matrix elements of the output and input states according to
\begin{equation}
\left[\mathcal E(\rho)\right]_{jk}=\sum_{m,n\in\mathbb{N}_0} \mathcal{E}^{mn}_{jk}\rho_{mn} ,
\end{equation}
where $\mathbb{N}_0\equiv \mathbb{N}\cup\{0\}$.

Below we show how to estimate process tensor elements $\mathcal{E}(|m\rangle \langle
n|)$ for $m,n$ over a finite domain. Because
\begin{equation}
 \bra{\alpha}(\ket{m}\bra{n})\ket{\alpha}=e^{-|\alpha|^2}\frac{\alpha^{n}{\bar{\alpha}}^m}{\sqrt{m!n!}}
\end{equation}
is in the Schwartz class, the Glauber-Sudarshan $P$ representation
\begin{equation}
 |m\rangle\langle n|=\int_{\mathds{C}} \mathrm{d}^{2}\alpha~
 P_{mn}(\alpha)|\alpha\rangle\langle \alpha|
\label{nm-}
\end{equation}
is guaranteed to exist for any  operator $|m\rangle\langle n|$
($m,n\in\mathbb{N}_0$)
\cite{Cahill2}.
The $P$ function is
\begin{equation}
\label{tempered-distribution}
 P_{mn}(\alpha)=(-1)^{m+n}\frac{ e^{|\alpha|^{2}}}{\sqrt{m!n!}} \partial_{\alpha}^m \partial_{\bar{\alpha}}^{n} \delta^{2}(\alpha)
\end{equation}
for $\partial_{\alpha}^m\coloneqq\partial^m/\partial\alpha^m$ and $\alpha$ and
its complex conjugate $\bar{\alpha}$ treated as independent variables, and $\delta^2(\alpha)\equiv\delta\bigl(\mathrm{Re}(\alpha)\bigr) \delta\bigl(\mathrm{Im}(\alpha)\bigr)$. By inserting representation~(\ref{nm-}) into Eq.~(\ref{superoperator}), and exploiting linearity of the
process, we obtain the process tensor
\begin{equation}
  \mathcal{E}^{mn}_{jk}=\int_{\mathds{C}} \mathrm{d}^{2}\alpha~ P_{mn}(\alpha)
  \langle j\rvert
\varrho_{\mathcal{E}}(\alpha)
\lvert k\rangle.
\label{sup}
\end{equation}
This expression can be simplified by using Eq.~(\ref{tempered-distribution}) and performing integration by parts:
\begin{align}
 \mathcal{E}^{mn}_{jk}&=\int_{\mathds{C}} \mathrm{d}^{2}\alpha~
 \frac{\delta^{2}(\alpha)}{\sqrt{m!n!}} \partial_{\alpha}^m
 \partial_{\bar{\alpha}}^{n}\bigl[ e^{|\alpha|^2}\langle j\rvert
\varrho_{\mbox{\tiny $\mathcal{E}$}}(\alpha)
\lvert k\rangle \bigr]\nonumber\\
&= \frac{1}{\sqrt{m!n!}} \partial_{\alpha}^m
 \partial_{\bar{\alpha}}^{n}\bigl[ e^{|\alpha|^2}\langle j\rvert
\varrho_{\mbox{\tiny $\mathcal{E}$}}(\alpha)
\lvert k\rangle \bigr]\Big|_{\alpha=0} .
\label{sup-op}
\end{align}

Thus we have eliminated the need to make use of the
Glauber-Sudarshan representation for quantum states. The process tensor
is found by taking partial derivatives (with respect to $\alpha$
and $\bar{\alpha}$) of the matrix elements of $\varrho_{\mathcal{E}}(\alpha)$, which are estimated from experimental data and evaluated at $\alpha=0$.

The mathematical procedure defined by Eq.~(\ref{sup-op}) is  simpler and computationally
faster (see Sec.~\ref{Sec:ExperimentalApplication}) than employing
Eq.~(\ref{sup}) with a regularized version of $P_{L,mn}(\alpha)$ replacing the
tempered distribution $P_{mn}(\alpha)$ described in
Refs.~\cite{LobinoQPT,LobinoQPT2}.
Equation~(\ref{sup-op}) has
been used to determine the fidelity of quantum teleportation of a single-rail optical qubit based on measurements performed on coherent states (see supplementary material in Ref.~\cite{PolzikNature2006}).

Generalization to the multi-mode case is straightforward.
In the $M$-mode case, let us introduce
the notation $|\mbox{\boldmath$n$}\rangle:=|n_1,n_2,\ldots,n_M\rangle$
(with $\mbox{\boldmath$n$}\in\mathbb{N}_0^M$)
for multi-mode Fock states and
$|\bm{\alpha}\rangle:=|\alpha_1,\alpha_2,\ldots,\alpha_M\rangle$
(with $\bm{\alpha}\in\mathds{C}^M$)
for multi-mode coherent states.
Then the matrix elements of the output and input states with respect
to the Fock basis are related to one another by the rank--{$\mathit 4^{M}$} tensor
\begin{equation}
\left[\mathcal E(\rho)\right]_{\mbox{\scriptsize\boldmath$j$}\mbox{\scriptsize\boldmath$k$}}\equiv\langle
\mbox{\boldmath$j$}|\mathcal E(\rho)|\mbox{\boldmath$k$}\rangle
=\sum_{\mbox{\scriptsize\boldmath$m$},\mbox{\scriptsize\boldmath$n$}\in\mathbb{N}_0^M}
\mathcal{E}^{\mbox{\scriptsize\boldmath$m$}\mbox{\scriptsize\boldmath$n$}}_{\mbox{\scriptsize\boldmath$j$}\mbox{\scriptsize\boldmath$k$}}\rho_{\mbox{\scriptsize\boldmath$m$}\mbox{\scriptsize\boldmath$n$}} ,
\end{equation}
where
\begin{equation}
\mathcal{E}^{\mbox{\scriptsize\boldmath$m$}\mbox{\scriptsize\boldmath$n$}}_{\mbox{\scriptsize\boldmath$j$}\mbox{\scriptsize\boldmath$k$}}:=\langle \mbox{\boldmath$j$}|
\mathcal E(|\mbox{\boldmath$m$}\rangle\langle\mbox{\boldmath$n$}|)|\mbox{\boldmath$k$}\rangle.
\end{equation}
Similarly to the single-mode case, we employ the Glauber-Sudarshan
$P$ representation for the multi-mode operator
$|\mbox{\boldmath$m$}\rangle\langle \mbox{\boldmath$n$}|$, with the overall $P$ function
being a product of the $P$ functions for the constituent modes:
\begin{equation}
 P_{\mbox{\scriptsize\boldmath$m$}\mbox{\scriptsize\boldmath$n$}}(\bm{\alpha}) = \prod_{s=1}^{M}\frac{e^{|\alpha_{s}|^2}(-1)^{m_{s}+n_{s}}}{\sqrt{m_{s}!n_{s}!}}
\partial_{\alpha_{s}}^{m_{s}} \partial_{\bar{\alpha}_{s}}^{n_{s}}\delta^{2}(\alpha_{s}).
\label{P-N}
\end{equation}
Multiple integration by parts yields
\begin{align}
\mathcal{E}^{\mbox{\scriptsize\boldmath$m$}\mbox{\scriptsize\boldmath$n$}}_{\mbox{\scriptsize\boldmath$j$}\mbox{\scriptsize\boldmath$k$}}&= \int_{\mathds{C}^M}\mathrm{d}^{2M}\bm{\alpha} \prod_{s=1}^{M} \frac{\delta^{2}(\alpha_{s})}{\sqrt{m_{s}!n_{s}!}} \partial_{\alpha_{s}}^{m_{s}} \partial_{\bar{\alpha_{s}}}^{n_{s}}
 \left[e^{|\alpha_{s}|^2}  \langle \mbox{\boldmath$j$}|
\varrho_{\mathcal{E}}(\bm{\alpha})
| \mbox{\boldmath$k$} \rangle\right]\;\nonumber\\
&=  \prod_{s=1}^{M} \frac{1}{\sqrt{m_{s}!n_{s}!}} \partial_{\alpha_{s}}^{m_{s}} \partial_{\bar{\alpha_{s}}}^{n_{s}}
 \left[e^{|\alpha_{s}|^2}  \langle \mbox{\boldmath$j$}|
\varrho_{\mathcal{E}}(\bm{\alpha})
| \mbox{\boldmath$k$} \rangle\right]\bigg|_{\alpha_s=0},
\label{E2}
\end{align}
where
\begin{equation}
\varrho_{\mathcal{E}}(\bm{\alpha})
 \equiv\mathcal E(|\bm{\alpha}
\rangle\langle\bm{\alpha}|).
\end{equation}
Equations (\ref{sup-op}) and (\ref{E2}) complete our coherent-state tomography
formalism and show that coherent states provide a complete set of probe
states for characterizing quantum optical processes, insofar as the expression
for $\varrho_{\mathcal{E}}(\bm{\alpha})$
completely determines the process tensor.

The above formalism is not restricted to trace-preserving quantum processes.
Indeed, trace preservation was not required in the derivation of our results.
Thus, our method is applicable to all quantum optical processes that are
mathematically described by completely-positive maps, but may be
trace-preserving, trace-reducing or even trace-increasing. Trace-nonpreserving
quantum processes are either {\em conditional processes} or part of
a larger process $\mathcal{E}=\mathcal{E}_1+\mathcal{E}_2$, which is
trace-preserving as a whole, but whose components $\mathcal{E}_1$ and
$\mathcal{E}_2$ may increase or decrease the trace, respectively.
A conditional process is a process that is conditioned on a certain probabilistic event; it may be heralded if the event is observed. One of the most notable examples of such a process is a probabilistic conditional-\textsc{not} gate
(\textsc{cnot}),
which
forms the basis for the Knill-Laflamme-Milburn linear-optical
quantum computing scheme~\cite{KLM}. Other examples are photon-addition
and photon-subtraction processes, whose superoperators
are derived in Sec.~\ref{Sec:Examples}.

In experimental csQPT, states $\varrho_{\mathcal{E}}(\alpha)$ are determined using homodyne tomography \cite{LvovskyRaymer}. It is important to remember, however, that this procedure reconstructs a density matrix normalized to unity trace: $\widetilde{\varrho}_{\mathcal{E}}(\alpha)=\varrho_{\mathcal{E}}(\alpha)/ \mathrm{Tr}\left[\varrho_{\mathcal{E}}(\alpha)  \right]$. When measuring non-trace-preserving processes, one must recover the trace information contained in  $\varrho_{\mathcal{E}}(\alpha)$. This is done by measuring the probability
$p_\alpha(\mathcal{E})=\mathrm{Tr}\left[\varrho_{\mathcal{E}}(\alpha)  \right]$
of the process heralding event for all $\alpha$'s for which the
measurements are performed. The state to be used in Eqs.~(\ref{sup-op}) and (\ref{E2})
in place of $\varrho_{\mathcal{E}}(\alpha)$ is then $\widetilde{\varrho}_{\mathcal{E}}(\alpha)\mathrm{Tr}\left[\varrho_{\mathcal{E}}(\alpha)  \right]$.

An interesting feature of Eqs.~(\ref{sup-op}) and (\ref{E2})
is that complete information about a quantum optical
process is contained in its action on an infinitesimally small compact
set of all probe coherent states in the immediate vicinity of the vacuum state.
From a mathematical point of view, this feature can be understood by realizing
that, for any $j,k\in \mathbb{N}_0$, the matrix element $\langle j\rvert
\varrho_{\mathcal{E}}(\alpha)
\lvert k\rangle $ is an {\em entire function} (see \ref{app:A}),
i.e.,
a complex-valued function in the variables $\alpha, \bar{\alpha}$ that is holomorphic over the whole complex plane,
and so is its product with the exponential $e^{|\alpha|^2}$.
Hence, each term $e^{|\alpha|^2}\langle j\rvert
\varrho_{\mathcal{E}}(\alpha)
\lvert k\rangle $ is infinitely differentiable over the whole complex plane
and is identical to its Taylor series expansion in any point of $\mathbb{C}$.
Moreover, Eq.~(\ref{sup-op}) implies that the process tensor is
determined by the corresponding Taylor coefficients at $\alpha=0$.
The same conclusion applies to the multi-mode case, in which we
deal with entire functions on $\mathbb{C}^M$.

\subsection{Energy cutoff and estimation of the error of approximation}
\label{SubSec:ErrorAnalysis}
As discussed in Sec.~1, the incompleteness of the information acquired in the experiment is accommodated in csQPT by evaluating the process tensor over a restricted finite-dimensional subspace $\widetilde{\mathcal{H}}$
of the Hilbert space $\mathcal{H}$  with a fixed maximum number $N$ of photons.
The incurred expense is that, through this reduced tomography, only approximate information about the
process will be inferred: for a given input state $\rho$, the predicted output is not $\mathcal{E}(\rho)$, but rather $\tilde{\mathcal{E}}(\tilde\rho)$, where 
\begin{equation}\label{rhotilde}
\tilde{\rho}=\frac{\widetilde{\Pi}\rho\widetilde{\Pi}}{\mathrm{Tr}[\rho\widetilde{\Pi}]} 
\end{equation}
is the trace-normalized projection of $\rho$ onto  $\mathcal{B}(\widetilde{\mathcal{H}})$ and
\begin{equation}\label{Etilde}
\tilde{\mathcal{E}}(\tilde\rho)=\widetilde{\Pi}{\mathcal{E}}(\tilde\rho)\widetilde{\Pi}
\end{equation}
is the predicted output of the reconstructed process for input state $\tilde\rho$. In Eqs.~\eqref{rhotilde} and \eqref{Etilde}, $\widetilde{\Pi}$ is the projection operator onto $\widetilde{\mathcal{H}}$.

If the input state $\rho$ is outside $\mathcal{B}(\widetilde{\mathcal{H}})$, the process output estimation error $\Vert\mathcal{E}(\rho)-\tilde{\mathcal{E}}(\tilde{\rho}\,)\Vert_1$ (where $\Vert\rho\Vert_1=\Tr\sqrt{\rho^\dagger\rho}$ is the trace norm) is generally unbounded. However, it is possible to bound the error for certain practically important classes of input states and processes.

\begin{figure}[tpb]
\centering
\epsfig{file=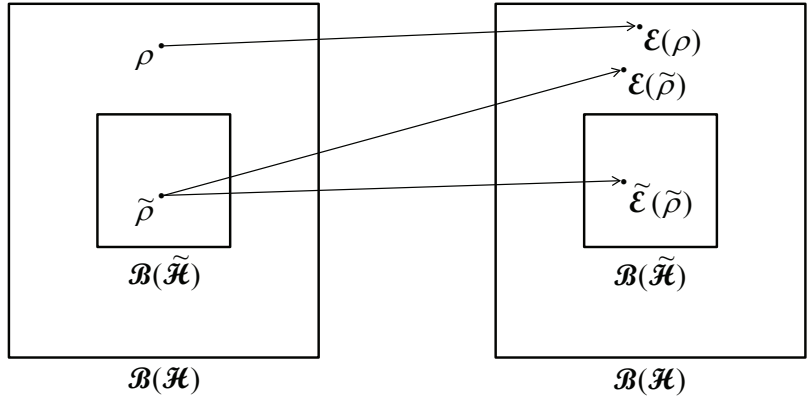,width=0.75\linewidth,clip=}
\vspace*{-9mm}
{\caption{Errors associated with photon number cutoff.
Restricting $\mathcal{B}(\mathcal{H})$ to $\mathcal{B}(\mathcal{\widetilde{H}})$ results in approximation $\widetilde{\rho}$ of the input state $\rho$.
If the error of this approximation $\Vert\rho-\widetilde{\rho}\,\Vert_1$ is known, the error of the images $\Vert\mathcal{E}(\rho)-\mathcal{E}(\widetilde{\rho}\,)\Vert_1$ can be estimated according to Eq.~(\ref{Eq:ErrorEstimation1}). However, the difference between $\mathcal{E}(\rho)$ and $\widetilde{\mathcal{E}}(\widetilde{\rho})$ in the cutoff space remains generally unknown.
\label{figure1}
}}
\end{figure}

For example, all linear-optical processes 
involving only linear-optical elements (interferometers, attenuators,
conditional measurements) do not generate additional photons, and thus map
$\mathcal{B}(\widetilde{\mathcal{H}})$ onto itself, so
$\tilde{\mathcal{E}}(\tilde{\rho})={\mathcal{E}}(\tilde{\rho})$. For such
processes, the error for a particular input $\rho$ can be estimated according to 
$\Vert\mathcal{E}(\rho)-\mathcal{E}(\widetilde{\rho})\Vert_1
\le\Vert\mathcal{E}\Vert\,\Vert\rho-\widetilde{\rho}\,\Vert_1$,
with the superoperator norm defined as $\Vert\mathcal{E}\Vert\equiv
\mbox{sup}\{\Vert\mathcal{E}(\hat {B})\Vert_1\,:\,\hat{ B}\in \mathcal{B}(\mathcal{H})\,,\,
\Vert \hat {B} \Vert_1\le 1\}$ \cite{Kitaev97}.
If the process is known to be trace-nonincreasing,
we have $\Vert\mathcal{E}\Vert\le 1$~\cite{NielsenChuang2000} so the error is bounded
from above by
\begin{equation}
\label{Eq:ErrorEstimation1}
\Vert\mathcal{E}(\rho)-\mathcal{E}(\widetilde{\rho}\,)\Vert_1
\le\Vert\rho-\widetilde{\rho}\,\Vert_1.\end{equation}
Note that the above result is not sufficient for evaluating the
error for a general process, because this error is given by the
deviation of $\mathcal{E}(\rho)$ from
$\widetilde{\mathcal{E}}(\widetilde{\rho})$ rather than from
 $\mathcal{E}(\widetilde{\rho})$
[Fig.~\ref{figure1}]. 


A further error bound can be obtained for the class of trace-preserving processes that do not increase
the mean energy, acting on a set of input states whose mean energy does not exceed a certain value \cite{D'Ariano2007}. We illustrate this for a single optical mode  $\hat{a}$
with frequency $\omega$ and Hamilton operator $\hat{H}=\omega(\hat{a}^{\dag}
\hat{a}+1/2)$ whose eigenvalues are denoted by $h_n=(n+1/2)\omega $. Suppose that the quantum states $\rho$ of
interest satisfy $\mathrm{Tr}[\rho\hat{H}]\le U$.
According to Ref.~\cite{D'Ariano2007}, if we choose the cutoff dimension
$\dim(\widetilde{\mathcal{H}})=N+1$
such that $U/h_{N+1}\le\gamma$ for some (small)
$\gamma>0$, the reconstructed process output errors are
bounded from above as
\begin{equation}
\left\Vert\mathcal{E}(\rho)-\frac{\mathcal{\widetilde{E}}(\widetilde{\rho})}{\Tr\mathcal{\widetilde{E}}(\widetilde{\rho})}\right\Vert_1 \le 2\epsilon,
\label{EE"}
\end{equation}
where
\begin{equation}
\epsilon = 2\sqrt{\gamma}+\gamma/(1-\gamma).\label{Epsilon=f(Gamma)}
\end{equation}

Conversely,
if we want to achieve a certain upper bound $\epsilon$ on the error of
approximation (which corresponds to a lower bound on the desired accuracy of
the process characterization), we first solve Eq.~(\ref{Epsilon=f(Gamma)}) for
$\gamma=\gamma(\epsilon)$, and then find the minimum $N_\gamma\in\mathbb{N}$
such that $U/h_{N_\gamma+1}\le\gamma $. Any cutoff dimension $N+1> N_\gamma $
is then sufficient for our purpose. For $\gamma\ll 1$, $\epsilon\approx
2\sqrt{\gamma}$, which yields
\begin{equation}
\label{error-scale}
\epsilon=O(1/\sqrt{N}).
\end{equation}
This implies that the error of
approximation scales as 
$1/\sqrt{N}$ with the cutoff dimension $N+1$.

For example, in order to achieve a 10\% error in \eeqref{EE"}, we need $\epsilon=0.05$ and thus $\gamma\approx 0.0006$. For the input mean energy bound corresponding to one photon ($U=3/2\omega$), the required cutoff is $N\approx U/\gamma\approx 250$. This calculation shows that the above error estimate is extremely conservative.

\section{Phase-invariant processes}
\label{Sec:ExperimentalApplication}

Many practically relevant processes, including the single-mode processes studied in Sec.~\ref{Sec:Examples}, exhibit phase invariance. If two input states are identical up to a shift by an optical phase $\phi$, the process outputs for these states differ by the same phase shift:
\begin{equation}\label{eq.shift}
   \mathcal{E}[e^{i\hat{n}\phi}\rho e^{-i\hat{n}\phi}]=e^{i\hat{n}\phi}\mathcal{E}(\rho) e^{-i\hat{n}\phi}.
\end{equation}
For such processes, it is convenient to express the probe coherent states in
polar coordinates:
$\ket\alpha=\ket {r e^{i\theta}}=e^{i\hat{n}\theta}\ket r$. Specifically, in these coordinates, we have \cite{Sudarshan}
\begin{equation}
\label{Ppolar}
 P_{mn}(r,\theta)=\frac{\sqrt{m!n!}}{(m+n)!} e^{r^2 +i\theta(n-m)} (-1)^{m+n}\frac{\mathrm{d}^{m+n}}{\mathrm{d} r^{m+n}}\delta(r),
\end{equation}
and accordingly
\begin{equation}
 \mathcal{E}^{mn}_{jk}
 =\frac{\sqrt{m!n!}}{(m+n)!}\frac{\mathrm{d}^{m+n}}{\mathrm{d} r^{m+n}}\left [\int_{0}^{2\pi} \frac{\text{d}\theta}{2\pi}
 e^{r^2 +i\theta(n-m)} \bra{j}\varrho_{\mbox{\tiny $\mathcal{E}$}}(r,\theta)
\ket{k} \right]\,\bigg|_{r=0}.
\label{sup-op-PolarForm}
\end{equation}
Hence Eq.~(\ref{eq.shift}) can be expressed as
\begin{equation}
 \bra{j}\mathcal{E}(\ket{\alpha}\bra{\alpha})\ket{k}
 =e^{i\theta(j-k)}\bra{j}\mathcal{E}(\ket{r}\bra{r})\ket{k},
\end{equation}
and the superoperator $\mathcal{E}$ [Eq.~(\ref{sup-op-PolarForm})] has the following explicit representation:
\begin{equation} \label{supopphaseinv}
\mathcal{E}^{ mn}_{jk}
 =\frac{\sqrt{m!n!}}{(m+n)!}\frac{\mathrm{d}^{m+n}}{\mathrm{d} r^{m+n}}\left[e^{r^2}\bra{j}\mathcal{E}(\ket{r}\bra{r})\ket{k}\right]\,\bigg|_{r=0} \delta_{m-j,n-k}.
\end{equation}

In experimental tomography of phase-invariant processes
\cite{LobinoQPT,LobinoQPT2},
%
it is sufficient to measure the process output for a discrete set of coherent states $\{\ket{r_i}\}$ on the real axis of the phase space.
The matrix elements of the output states can then be interpolated as polynomial functions
\begin{equation}
\label{poly}
\bra{j}\mathcal{E}(\ket{r}\bra{r})\ket{k}=\sum_{l=0}^{Q}C_l(j,k)r^{l},
\end{equation}
where $Q$ is the degree of the polynomial (which depends on the dimension of
the truncated Hilbert space) and $C_l(j,k)$ are its coefficients. Furthermore,
from Eq.~(\ref{ConvergentSeries}) together with  Eq.~(\ref{supopphaseinv}),
it follows that, for phase-symmetric processes, when $j-k$ is even or odd,
 $\bra{j}\mathcal{E}(\ket{r}\bra{r})\ket{k}$
 and
 its analytic extension
 to negative values of $r$
 are
 even
or
odd functions of $r$, respectively. By taking into account the symmetric or antisymmetric
property of this function, we have additional information to be
used in the interpolation procedure; the constructed polynomial has
to contain only even or odd powers of $r$, respectively.
In this way the precision of process estimation from the experimental data is
substantially increased.

With the knowledge of the coefficients $C_l(j,k)$, \eeqref{supopphaseinv} is further simplified to:
\begin{align}
\label{sup-op2}
\mathcal{E}^{mn}_{jk} & =\frac{\sqrt{m!n!}}{(m+n)!}\frac{\mathrm{d}^{m+n}}{\mathrm{d} r^{m+n}}\left[\sum_{s=0}^{\infty}\frac{r^{2s}}{s!}
\sum_{l=0}^{Q} C_l(j,k) r^l\right]\Bigg|_{r=0} \delta_{m-j,n-k}\nonumber \\
& =\frac{\sqrt{m!n!}}{(m+n)!}\sum_{s=0}^{\infty}\sum_{l=0}^{Q}\frac{\delta_{m+n,2s+l}(m+n)!}{s!}C_l(j,k) \delta_{m-j,n-k}\nonumber \\
& =\sqrt{m!n!}\sum_{s=0}^{\lfloor(m+n)\rfloor/2}\frac{C_{m+n-2s}(j,k)}{s!}\delta_{m-j,n-k}\,.
\end{align}
The last result is significant in that one can obtain the process
tensor directly from the experimentally reconstructed output states through simple summation.
Moreover, if the dimension of the truncated Hilbert space
is $d=N+1$, from Eq.~(\ref{sup-op2}) it follows that only terms of power
$l\le 2N$ of the interpolation polynomial (\ref{poly}) contribute to the
process tensor. We have tested this procedure on
experimental data \cite{LobinoQPT2} and calculated the process
tensor in a few microseconds, which is
a dramatic improvement in comparison to
several hours required for the original procedure \cite{LobinoQPT,LobinoQPT2}.

\section{Examples: superoperators of important  quantum optical processes}
\label{Sec:Examples}

In this section, we illustrate our new method by applying it to some fundamental quantum optical processes, whose effects on coherent states are known. Specifically, using Eqs.~(\ref{sup-op}) or (\ref{E2}), we analytically derive corresponding
superoperator tensors $\mathcal{E}^{mn}_{jk}$ in the Fock basis. The results are summarized in
Table~\ref{p-tab}.

\subsection{Identity}

For the  identity process ($\mathcal{E}_{\text{id}}$), $\varrho_{\mathcal{E}_{\text{id}}}(\alpha)=|\alpha\rangle\langle\alpha|$, the matrix
elements of the output states are
\begin{equation}
\bra{j}\varrho_{\mathcal{E}_{\text{id}}}(\alpha)
\ket{k}=e^{-|\alpha|^2}\frac{\alpha^j{\bar{\alpha}}^{k}}{\sqrt{j!k!}}\;.
\label{identity-matrixelements}
\end{equation}
Inserting these elements into
Eq.~(\ref{sup-op}) yields $\mathcal{E}^{mn}_{jk}=\delta_{mj}\delta_{nk}$, as expected.

\subsection{Attenuation and lossy channel}

For attenuation of light fields ($\mathcal{E}_{\text{att}}$),
the process's effect on single-mode coherent states is given
by $\varrho_{\mathcal{E}_{\text{att}}}(\alpha)=|\eta\alpha\rangle\langle\eta\alpha|$, where $0\leq\eta<1$.
The matrix elements in the Fock basis are
\begin{equation}
 \bra{j} \varrho_{\mbox{\tiny
    $\mathcal{E}_{\mbox{\tiny att}}$}}(\alpha)\ket{k}
=e^{-\eta^{2}|\alpha|^2}\frac{\eta^{j+k}\alpha^j{\bar{\alpha}}^{k}}{\sqrt{j!k!}}\;.
\end{equation}
From
Eq.~(\ref{sup-op}), we obtain
\begin{align}
\mathcal{E}^{mn}_{jk}&=\frac{\eta^{j+k}}{\sqrt{m!n!j!k!}}\partial^{m}_{\alpha}\partial^{n}_{\bar{\alpha}}\left[e^{|\alpha|^{2}(1-\eta^{2})}\alpha^{j}{\bar{\alpha}}^{k}\right]\Big|_{\alpha,\bar{\alpha}=0}\nonumber\\
&=\frac{\eta^{j+k}}{\sqrt{m!n!j!k!}}\partial^{m}_{\alpha}\partial^{n}_{\bar{\alpha}}
\sum_{l=0}^{\infty}\frac{(1-\eta^{2})^{l}\alpha^{j+l}{\bar{\alpha}}^{k+l}}{l!}\Big|_{\alpha,\bar{\alpha}=0}\nonumber\\
&=\sqrt{\frac{m!n!}{j!k!}}\frac{\eta^{j+k}(1-\eta^{2})^{m-j}}{(m-j)!}\delta_{m-j,n-k}\,,\label{Eq:TensorForAttenuation}
\end{align}
which depends explicitly on $\eta$.

\subsection{Photon subtraction and addition}

Photon subtraction is defined as a process that removes a single photon from the
light field, whereas photon addition adds a single photon. Photon subtraction has
been used by Ourjoumtsev et al. \cite{GrangierScience2006}
to generate optical Schr{\"o}dinger kittens (coherent superpositions of low-amplitude coherent states) from squeezed vacuum states
for the purpose of quantum information processing.
Single-photon-added coherent states can be
regarded as the result of the most elementary amplification process
of classical light fields by a single quantum of excitation; being
intermediate between single-photon Fock states (fully
quantum-mechanical) and coherent (classical) ones,
these states have been demonstrated to be suited for
the study of smooth transition between the particle-like
and the wavelike behavior of light~\cite{BelliniScience2006}.

\begin{figure}[tpb]
\centering
\epsfig{file=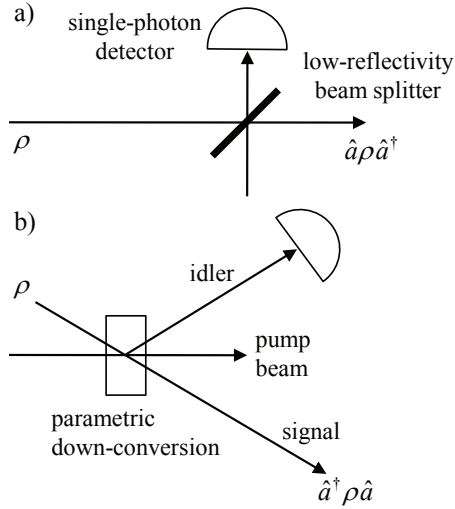,width=0.4\linewidth,clip=}
{\caption{Experimental realizations of (a) photon subtraction and (b) photon
    addition. The process is heralded by single-photon detection events.
\label{figure2}
}}
\end{figure}

Here we discuss idealized single-mode photon subtraction and photon addition.
Both processes are non-trace-preserving. For example, photon subtraction
can be approximately realized \cite{GrangierScience2006} by a highly-transmissive beam splitter, whose
reflected mode is directed to a detector
and whose transmitted mode constitutes
the output, respectively, as illustrated in Fig.~\ref{figure2}a.
Any click in a detector implies extraction of photon(s) from the input mode by the beam splitter. As the beam splitter has low reflectivity, here single-photon extraction events are more likely than multi-photon events.
%
An approximate experimental realization of photon addition is illustrated in
Fig.~\ref{figure2}b. The input quantum state $\rho$
enters the signal channel of a parametric down-conversion setup.
Provided that detector dark counts are neglected,
a photon detection in the idler mode heralds photon addition
to the signal mode, which contains the output state of the process.

The effect of photon subtraction ($\mathcal{E}_{\mbox{\scriptsize sub}}$)
and addition ($\mathcal{E}_{\mbox{\scriptsize add}}$)
on coherent states is given by
$\varrho_{\mathcal{E}_{\text{sub}}}(\alpha)=\hat{a}|\alpha\rangle\langle\alpha|\hat{a}
^{\dagger}$ and $\varrho_{\mathcal{E}_{\text{add}}}(\alpha)=\hat{a}
^{\dagger}|\alpha\rangle\langle\alpha|\hat{a}$, respectively,
where $\hat{a}$ and $\hat{a}^{\dagger}$ are the photon annihilation
and photon creation operators of a single mode, respectively.
The matrix elements of the output states in the Fock basis are
\begin{align}
\bra{j} \varrho_{\mathcal{E}_{\text{sub}}}(\alpha)\ket{k}
&=e^{-|\alpha|^2}\frac{\alpha^{j+1}{\bar{\alpha}}^{k+1}}{\sqrt{j!k!}},\\
\bra{j} \varrho_{\mathcal{E}_{\text{add}}}(\alpha)\ket{k}
&=
e^{-|\alpha|^2}\sqrt{kj}\frac{\alpha^{j-1}{\bar{\alpha}}^{k-1}}{\sqrt{(j-1)!(k-1)!}}\,.
\end{align}
The process tensor is found to be
\begin{equation}
\mathcal{E}^{ mn}_{jk}=
\left\{  \begin{array}{ll}
\sqrt{(j+1)(k+1)}\delta_{m,j+1}\delta_{n,k+1},& \mbox{for photon subtraction,}
\\
\sqrt{kj}\delta_{m,j-1}\delta_{n,k-1},& \mbox{for photon addition,}
\\
\end{array}  \right.
\end{equation}
where we have employed Eq.~(\ref{sup-op}).

\subsection{Schr{\"o}dinger cat generation}

The unitary evolution according to $\hat{U}_{\text{Kerr}}(\chi)\equiv
\exp\left[-i\chi\left(\hat{a}
^{\dagger}\hat{a}\right)^2\right]$ for $\chi=\pi/2$, if applied to coherent states, generates
Schr{\"o}dinger cat states (hereafter denoted as $\mathcal{E}_{\text{cat}}$) \cite{M86,YS86}
\begin{align}
\varrho_{\mathcal{E}_{\text{cat}}}(\alpha)&= \hat{U}_{\text{Kerr}}(\frac{\pi}{2})\ketbra\alpha\alpha\hat{U}^\dagger_{\text{Kerr}}(\frac{\pi}{2})
\nonumber\\
&=\frac{1}{2}(\ket{\alpha}+i\ket{-\alpha})(\bra{\alpha}-i\bra{-\alpha}),
\end{align}
with matrix elements
\begin{equation}
  \bra{j}
\varrho_{\mbox{\tiny
    $\mathcal{E}_{\text{cat}}$}}(\alpha)
\ket{k}=\frac{e^{-|\alpha|^2}\alpha^j{\bar{\alpha}}^{k}}{2\sqrt{j!k!}}\left[1+(-1)^{j+k}+i(-1)^j-i(-1)^k\right].
\end{equation}
The superoperator tensor for this non-Gaussian unitary process
obtained via
Eq.~(\ref{sup-op}) is
\begin{equation}
\mathcal{E}^{mn}_{jk}=e^{-i\frac{\pi}{2}(j^2-k^2)}\delta_{mj}\delta_{nk}\,.
\end{equation}
Interestingly, this process does not change the total particle number of any input state.

\subsection{Beam splitter}

Now let us consider the beam splitter as an example of a two-mode
process. The unitary beam splitter transformation is given by \cite{YMK86}
\begin{equation}
\hat{B}(\Theta)=e^{\frac{\Theta}{2}(\hat{a}^{\dagger}_2\hat{a}_1-\hat{a}^{\dagger}_1\hat{a}_2)},
\end{equation}
where $\Theta$ is the parameter identifying how the beam splitter transmits or reflects beams. Specifically, its action on coherent state inputs $\ket{\alpha_1}$
and $\ket{\alpha_2}$ is given as
\begin{align}
\varrho_{\mbox{\tiny
    $\mathcal{E}_B$}}(\alpha_1,\alpha_2)=& \mathcal{E}_B( |\alpha_1,\alpha_2\rangle\langle\alpha_1,\alpha_2|)\nonumber\\
=&
\hat{B}^{\dagger}(\Theta)( |\alpha_1,\alpha_2\rangle\langle\alpha_1,\alpha_2|)\hat{B}(\Theta)\nonumber\\
=&
|T\alpha_1-R\alpha_2,R\alpha_1+T\alpha_2\rangle\langle T\alpha_1-R\alpha_2,R\alpha_1+T\alpha_2|,
\end{align}
with $T\equiv\cos(\Theta/2)$ and $R\equiv\sin(-\Theta/2)$ being the transmissivity and reflectivity, respectively. By knowing the effect of the process on two-mode coherent states,
we can calculate the corresponding tensor using
Eq.~(\ref{E2}), which yields
\begin{align}
\mathcal{E}^{m_1m_2n_1n_2}_{j_1j_2k_1k_2}=& \sqrt{\frac{m_1!m_2!n_1!n_2!}{j_1!j_2!k_1!k_2!}}
\sum_{p=0}^{j_{1}}\sum_{q=0}^{k_1}\binom{j_1}{p}\binom{j_2}{m_1-p}\binom{k_1}{q}\nonumber \\
& \times \binom{k_2}{n_1-q} T^{2p+2q+j_2+k_2-m_1-n_1}\nonumber \\
& \times (-1)^{j_1+k_1-p-q} R^{j_1+k_1+m_1+n_1-2p-2q}\nonumber \\
& \times \delta_{m_1+m_2,j_1+j_2}\delta_{n_1+n_2,k_1+k_2},
\end{align}
as an explicit function of $T$ and $R$.

\subsection{Parametric down-conversion}

Another two-mode process of interest is parametric
down-conversion (PDC). In PDC, a crystal with an appreciably large second-order non-linearity is pumped by a laser field. Each of the pump photons
can spontaneously decay into a pair of identical (degenerate
PDC) or nonidentical photons (nondegenerate PDC). Here we consider
a nondegenerate PDC process $\mathcal{E}_{\text{PDC}}$ induced by the transformation
\cite{YMK86}
 \begin{equation}
 \hat{S}_2(r)=e^{r(\hat{a}_1\hat{a}_2-\hat{a}_1^{\dagger}\hat{a}_2^{\dagger})}.
\end{equation}
The effect of this unitary process on a two-mode coherent state is given by
\begin{align}
 \varrho_{\mathcal{E}_{\text{PDC}}}
(\alpha_1,\alpha_2)&=\mathcal{E}_{\mbox{\tiny PDC}}(\ket{\alpha_1,\alpha_2}\bra{\alpha_1,\alpha_2})\nonumber\\
&=\hat{S}_2(r)\ket{\alpha_1,\alpha_2}\bra{\alpha_1,\alpha_2}\hat{S}_2^{\dagger}(r).
\end{align}
In
\ref{app:B},
we derive the process tensor in the Fock
basis. The result can be expressed as:
\begin{align}
 \mathcal{E}^{m_1m_2n_1n_2}_{j_1j_2k_1k_2}
=&\sqrt{\frac{n_1!m_1!m_2!n_2!}{j_1!k_1!k_2!j_2!}}
\frac{(\tanh r )^{m_1+n_1-j_1-k_1}}{(m_1-j_1)!\ (n_1-k_1)!\
(\cosh r)^{j_2+k_2-j_1-k_1+2}}\nonumber \\
&\times\, _2F_{1}\left(-j_1,m_2+1;m_1-j_1+1;\tanh^2 r\right)\nonumber \\
&\times\, _2F_{1}\left(-k_1,n_2+1;n_1-k_1+1;\tanh^2 r\right)\nonumber \\
&\times \delta_{m_2-m_1,j_2-j_1}\  \delta_{n_2-n_1,k_2-k_1}\ ,
\label{Eq:PDC-Tensor}
\end{align}
with
\begin{equation}
_2F_1(\alpha,\beta;\gamma;z)\label{Eq:DefHypergeometricFunction}
:= 1+\sum_{n=1}^\infty \frac{(\alpha)_n(\beta)_n}
{(\gamma)_n}\frac{z^n}{n!} \,,
\end{equation}
the hypergeometric function, $(x)_n:=\Gamma(x+n)/\Gamma(x)$ the Pochhammer symbol
and $\Gamma(\cdot)$ the Gamma function \cite{Abram}.

\begin{table}[tp]
\caption{Process tensor
  $\mathcal{E}_{\mbox{\scriptsize\boldmath$j$}\mbox{\scriptsize\boldmath$k$}}^{\mbox{\scriptsize\boldmath$m$}\mbox{\scriptsize\boldmath$n$}}$
for some quantum optical processes.  }
\vspace{1mm}
\begin{indented}
\item[]\hspace*{-2cm}
\begin{tabular}{@{}lllll}
\br
Operation $\mathcal{E}$ &  $\varrho_{\mbox{\tiny
    $\mathcal{E}$}}(\alpha)$   &Process tensor $\mathcal{E}_{\mbox{\scriptsize\boldmath$j$}\mbox{\scriptsize\boldmath$k$}}^{\mbox{\scriptsize\boldmath$m$}\mbox{\scriptsize\boldmath$n$}}$ \\
\mr
Identity  ($\mathcal{E}_{\mbox{\scriptsize id}}$) & $ \ketbra\alpha\alpha  $ &  $\delta_{mj}\delta_{nk}$  \\ &&\\
Attenuation  ($\mathcal{E}_{\mbox{\scriptsize att}}$) & $
|\eta\alpha\rangle\langle \eta\alpha| $ &
$\sqrt{\frac{m!n!}{j!k!}}\frac{\eta^{j+k}(1-\eta^{2})^{m-j}}{(m-j)!}\delta_{m-j,n-k} $
\\&&\\
Photon addition    ($\mathcal{E}_{\mbox{\scriptsize add}}$) & $ \hat{a}^{\dag}|\alpha\rangle\langle \alpha|\hat{a}  $   &  $\sqrt{kj}\delta_{m,j-1}\delta_{n,k-1}$  \\&&\\
Photon subtraction ($\mathcal{E}_{\mbox{\scriptsize sub}}$) &
$ \hat{a}|\alpha\rangle\langle \alpha|\hat{a}^{\dagger}  $  &  $ \sqrt{(j+1)(k+1)}\delta_{m,j+1}\delta_{n,k+1}$  \\&&\\
Cat generation ($\mathcal{E}_{\mbox{\scriptsize
    cat}}$) & $ \frac{1}{2}(\ket{\alpha}+i\ket{-\alpha})$
&  $e^{-i\frac{\pi}{2}(j^2-k^2)}\delta_{mj}\delta_{nk}$  \\
&$\quad\times(\bra{\alpha}-i\bra{-\alpha}) $ & \\  \\&&\\
Beam splitter ($\mathcal{E}_{B}$) &
$|T\alpha_1-R\alpha_2,R\alpha_1+T\alpha_2\rangle  $  &
$\sqrt{\frac{m_1!m_2!n_1!n_2!}{j_1!j_2!k_1!k_2!}}
\sum_{p=0}^{j_1}\sum_{q=0}^{k_1}(-1)^{j_1+k_1-p-q}$\\
&  $\times\langle T\alpha_1-R\alpha_2,R\alpha_1+T\alpha_2|$ &
 $\times\binom{j_1}{p}\binom{j_2}{m_1-p}\binom{k_1}{q}
 \binom{k_2}{n_1-q}$\\ &     &$\times
  T^{2p+2q+j_2+k_2-m_1-n_1}$ \\ &  &$\times R^{j_1+k_1+m_1+n_1-2p-2q}$ \\  & &
 $\times \delta_{m_1+m_2,j_1+j_2}\delta_{n_1+n_2,k_1+k_2}$
\\&&\\
\\&&\\
Parametric down-  &
$e^{r(\hat{a}_1\hat{a}_2-\hat{a}_1^{\dagger}\hat{a}_2^{\dagger})}
\ket{\alpha_1,\alpha_2}
$ & $\sqrt{\frac{m_1!m_2!n_1!n_2!}{j_1!j_2!k_1!k_2!}}$\\ conversion   ($\mathcal{E}_{\mbox{\scriptsize PDC}}$)
&
$\times\bra{\alpha_1,\alpha_2}e^{r(\hat{a}_1^{\dagger}\hat{a}_2^{\dagger}-\hat{a}_1\hat{a}_2)}$
& $\times\frac{(\tanh r)^{m_1+n_1-j_1-k_1}}{(m_1-j_1)!\ (n_1-k_1)!\
(\cosh r)^{j_2+k_2-j_1-k_1+2}}$ \\
&  &$\times\, _2F_{1}\left(-j_1,m_2+1;m_1-j_1+1;\tanh^2 r\right)$\\
&  &$\times\,_2F_{1}\left(-k_1,n_2+1;n_1-k_1+1;\tanh^2 r\right)$ \\
&  &$\times\,\delta_{m_2-m_1,j_2-j_1} \delta_{n_2-n_1,k_2-k_1}$\\
\br
\end{tabular}
\end{indented}
\label{p-tab}
\end{table}

\section{Conclusions}
\label{conc}


Coherent states are easily generated probe states  for
tomography of unknown quantum-optical processes.
Here, we have presented a new, more efficient data processing technique for estimating a quantum process from similar experimental procedure of Ref.~\cite{LobinoQPT}.
The
original formulation
%
was
based on regularization and filtering of the Glauber-Sudarshan representations for
quantum states, which are cumbersome to implement numerically.
Furthermore, Ref.~\cite{LobinoQPT} introduces additional errors associated with regularization of the $P$ function.
%
%
%
%
In contrast, our new method to determine the process superoperator
[Eq.~(\ref{sup-op}) or Eq.~(\ref{E2})]
is mathematically simpler, computationally faster and unique up to the choice of the energy cutoff. Moreover, we presented straightforward generalizations of
coherent state quantum process tomography
to multi-mode and non-trace-preserving conditional processes.

We have illustrated the new framework
through
several examples (summarized in
Table~\ref{p-tab}).
%
We have shown that it is straightforward to derive analytically
exact and unique closed-form expressions for the superoperators for quantum optical
processes whose effect on coherent states is known. For
phase-invariant
unknown processes, the formula to find the process tensor reduces
to a simple summation of coefficients of a polynomial obtained
from the experimentally reconstructed output states via interpolation.

An interesting consequence implied by our formulation [in particular, Eqs.~(\ref{sup-op}) and (\ref{E2})]
is
that complete information
about a quantum optical process is entirely captured by its effect
on a compact set of all coherent states $\ket{\alpha}$ in the immediate
vicinity of the vacuum state.
This is due to the entireness property of the image of processes on coherent states.
It thus appears sufficient to
perform
tomography
experiments only for a range of coherent states whose
mean photon number is much smaller than that required for the method of
Ref.~\cite{LobinoQPT} (see the suppl.\ material therein).
However, coherent state quantum process tomography
relies on the ability to approximately determine all the derivatives
of a function which is obtained by interpolation from measured experimental
data. Minimization of errors associated with this calculation imposes a lower
bound on the phase space region over which the measurements need to be
performed. For the time being,
we have provided an evaluation of the error in the process estimation
by introducing a truncation of the Fock space. For the class of processes
respecting a certain energy constraint (which includes all processes that
do not amplify the energy), we have determined
(i) the cutoff dimension that is sufficient in order to achieve a certain degree of
approximation accuracy, as well as (ii) the upper bound on the error of estimation
for a given cutoff dimension.

\vspace*{3mm}
\noindent{\bf Acknowledgments:}

We acknowledge financial support by NSERC, $i$CORE, MITACS, QuantumWorks and General
Dynamics Canada. AIL is a CIFAR Scholar, and BCS is a CIFAR Fellow. We would
also like to thank Connor Kupchak for helpful discussions.

\appendix

\section{ Proof that {\boldmath{$\bra{j}\varrho_{\mathcal{E}}(\alpha)\ket{k}$}} is an entire function}
\label{app:A}

According to Eq.~(\ref{sup-op}), by knowing the complex-valued function
$\bra{j}\varrho_{\mathcal{E}}(\alpha)\ket{k}$ (of the variable $ \alpha$)
for any $j$ and $k$, one can determine the process tensor
$\mbox{\boldmath$E$}^{mn}_{jk}$.
Here we show that this function is an {\em entire function}
so it can be represented as a power
series that converges uniformly on any compact domain.

As a
completely-positive
quantum operation, $\mathcal{E}$ possesses a  Kraus decomposition
$\mathcal{E}(\rho) = \sum_{i=1}^{L}\hat{K}_{i}^{}
\, \rho \,\hat{K}_{i}^{\dagger}$, where
$L \leq \dim(\mathcal{H})^2$
and $\hat{K}_{i}$ are some Kraus operators on $\mathcal{H}$
(whose explicit form is not needed for our purpose).
Hence we can rewrite the matrix elements of the output state as
\begin{align}
\label{invpro}
 \bra{j}\varrho_{\mathcal{E}}(\alpha)\ket{k}&=\sum_{i=1}^{L} \bra{j}\hat{K}_{i}^{} \,
\ket{\alpha}\bra{\alpha} \,\hat{K}_{i}^{\dagger}\ket{k}\; \nonumber\\
&=\sum_{i=1}^{L}\bra{\alpha} \,
\hat{K}_{i}^{\dagger}\ket{k}\bra{j}\hat{K}_{i}^{} \, \ket{\alpha}\nonumber \\
&=\bra{\alpha} \mathcal{E}_{*}(\ket{k}\bra{j})  \ket{\alpha},
\end{align}
where
\begin{equation}
\mathcal{E}_*:\mathcal{B}(\mathcal{H})\rightarrow\mathcal{B}(\mathcal{H})
\,,\,\hat{B}\mapsto \sum_{i=1}^{L}\hat{K}_{i}^{\dagger}
\, \hat{B} \,\hat{K}_{i},
\end{equation}
is the dual or adjoint map \cite{Werner:08}.
The complex-valued function
$\bra{\alpha}\hat{A}\ket{\alpha}$ (referred to as Husimi
function if $\hat{A}$ is a density operator)---where
$\hat{A}$ is any bounded operator on $\mathcal{H}$---is an entire function of the two
variables $\alpha$ and $\bar{\alpha}$ \cite{Cahill1,MandelWolf}.
Hence the right hand
side of Eq.~(\ref{invpro}) implies that the function
$\bra{j}\varrho_{\mathcal{E}}(\alpha)\ket{k}$ is an entire function.
By representing the coherent states in
Eq.~(\ref{invpro})
in the Fock basis and using Eq.~(\ref{superoperator}), we obtain
\begin{equation}
 \bra{j}\varrho_{\mathcal{E}}(\alpha)\ket{k}
=e^{-|\alpha|^2}\sum_{n=0}^{\infty}\sum_{m=0}^{\infty}\frac{\alpha^n
\bar{\alpha}^m}{\sqrt{n!m!}}\mathcal{E}^{nm}_{jk}\, ,\label{ConvergentSeries}
\end{equation}
which is
a power series of the complex variables $\alpha$ and $\bar{\alpha}$, hence convergent everywhere \cite{Cahill1,MandelWolf}.

\section{Process tensor for parametric down-conversion}
\label{app:B}

To obtain the Fock representation of the PDC process,
we
first find the matrix elements of the output states in the Fock basis:
\begin{align}
 \bra{j_1,j_2}\varrho_{\mathcal{E}_{\text{PDC}}}(\alpha_1,\alpha_2)\ket{k_1,k_2}&=\bra{j_1,j_2}\hat{S}_2(r)\ket{\alpha_1,\alpha_2}\bra{\alpha_1,\alpha_2}\hat{S}_2^{\dagger}(r)\ket{k_1,k_2}\nonumber
\\
&=I\times \bar{J},
\end{align}
where $I:=\bra{\alpha_1,\alpha_2}\hat{S}_2^{\dagger}(r)\ket{k_1,k_2}$ and
$J:=\bra{\alpha_1,\alpha_2}\hat{S}_2^{\dagger}(r)\ket{j_1,j_2}$.
Employing the relations
\begin{eqnarray}
 \hat{S}_2^{\dagger}(r)\ket{0,0}&=&\frac{1}{\cosh r}\sum_{l=0}^{\infty}(\tanh r)^{l}\ket{l,l},\\
 \hat{S}_2^{\dagger}(r) \hat{a}_1
\hat{S}_2(r)&=&\hat{a}_1\cosh r-\hat{a}_2^{\dagger}\sinh r, \\
\hat{S}_2^{\dagger}(r) \hat{a}_2
\hat{S}_2(r)&=&\hat{a}_2\cosh r-\hat{a}_1^{\dagger}\sinh r,
\end{eqnarray}
and
the binomial expansion, we have:
\begin{align}
 I=&\bra{\alpha_1,\alpha_2}\hat{S}_2^{\dagger}(r)\frac{(\hat{a}_1^{\dagger})^{k_1}}{\sqrt{k_1!}}\frac{(\hat{a}_2^{\dagger})^{k_2}}{\sqrt{k_2!}}\hat{S}_2(r)\hat{S}_2^{\dagger}(r)\ket{0,0}\nonumber\\
=& \frac{1}{\cosh r\sqrt{k_1!k_2!}}\sum_{l=0}^{\infty}(\tanh r)^{l}\bra{\alpha_1\alpha_2}\sum_{p=0}^{k_1}\binom{k_1}{p}(\hat{a}_1^{\dagger}\cosh r)^{k_1-p}(-\hat{a}_2\sinh r)^{p}\nonumber\\
&\times
\sum_{q=0}^{k_2}\binom{k_2}{q}(\hat{a}_2^{\dagger}\cosh r)^{k_2-q}(-\hat{a}_1\sinh r)^{q}\ket{l,l}.\label{I2}
\end{align}
Using  $\hat{a}^x\ket{l}=\sqrt{l!/(l-x)!}\ket{l-x}$ and
$(\hat{a}^{\dagger})^{y}\ket{l}=\sqrt{(l+y)!/l!}\ket{l+y}$ we obtain:
\begin{align}
 I=&\frac{1}{\cosh r\sqrt{k_1!k_2!}}\sum_{l=0}^{\infty}(\tanh r)^{l}\sum_{p=0}^{k_1}\binom{k_1}{p}(\cosh r)^{k_1-p}(-\sinh r)^{p}
\nonumber \\
&\times\sum_{q=0}^{k_2}\binom{k_2}{q}(\cosh r)^{k_2-q}(-\sinh r)^{q}
e^{-|\alpha_1|^2/2-|\alpha_2|^2/2}\nonumber \\
&\times\bar{\alpha}_1^{l+k_1-q-p}\bar{\alpha}_2^{l+k_2-q-p}\frac{(l+k_2-q)!}{(l-q)!(l+k_2-q-p)!}\;.
\end{align}
 From the symmetry between $I$ and $J$, and by replacing $k_1$ and $k_2$ by
 $j_1$ and
$j_2$, respectively, we also find:
\begin{align}
\bar{J}=&\frac{1}{\cosh r\sqrt{j_1!j_2!}}\sum_{l'=0}^{\infty}(\tanh r)^{l'}\sum_{u=0}^{j_1}\binom{j_1}{u}(\cosh r)^{j_1-u}(-\sinh r)^{u}
\nonumber \\
&\times\sum_{v=0}^{j_2}\binom{j_2}{v}(\cosh r)^{j_2-v}(-\sinh r)^{v}\
e^{-|\alpha_1|^2/2-|\alpha_2|^2/2}\nonumber \\
&\times \alpha_1^{l'+j_1-u-v}\alpha_2^{l'+j_2-u-v}\frac{(l'+j_2-v)!}{(l'-v)!\
(l'+j_2-u-v)!}\; .
\end{align}
The Fock representation of the superoperator for the PDC process is then given
by
\begin{align}
\hspace*{-2cm} \mathcal{E}^{m_1m_2n_1n_2}_{j_1j_2k_1k_2}=&\frac{1}{\sqrt{m_1!m_2!n_1!n_2!}}\partial^{m_1}_{\alpha_1}
\partial^{n_1}_{\bar{\alpha}_1}\partial^{m_2}_{\alpha_2}
\partial^{n_2}_{\bar{\alpha}_2}\left(e^{|\alpha_1|^2+|\alpha_2|^2}
I \times \bar{J}\right)\bigg|_{\alpha_1,\alpha_2=0}\nonumber \\
=&\sqrt{\frac{n_1!m_1!}{n_2!m_2!}}\
\frac{1}{\cosh r^2\sqrt{j_1!j_2!k_1!k_2!}}\sum_{p=0}^{k_1}\binom{k_1}{p}(\cosh r)^{k_1-p}(-\sinh r)^{p}\nonumber\\
&\times (\tanh r)^{n_1-k_1+p}\frac{(n_1-k_1+k_2+p)!}{(n_1-k_1+p)!}\
\delta_{n_1-n_2,k_2-k_2}\nonumber\\
&\times
\sum_{q=0}^{k_2}\binom{k_2}{q}(\cosh r)^{k_2-q}(-\sinh r\tanh r)^{q}\nonumber \\
 &\times\sum_{u=0}^{j_1}\binom{j_1}{u}(\cosh r)^{j_1-u}
(-\sinh r)^{u}(\tanh r)^{m_1-j_1+u}\frac{(m_1-j_1+j_2+u)!}{(m_1-j_1+u)!}\nonumber \\
&\times
\delta_{m_2-m_1,j_2-j_1}\sum_{v=0}^{j_2}\binom{j_2}{v}(\cosh r)^{j_2-v}(-\sinh r\tanh r)^{v}\nonumber \\
=&\sqrt{\frac{n_1!m_1!k_1!j_1!}{n_2!m_2!k_2!j_2!}}\
\frac{(\tanh r)^{n_1+m_1}}{(\cosh r)^{k_2+j_2+2}}\ \nonumber \\
&\times \sum_{p=0}^{k_1}
\sum_{u=0}^{j_1}
\frac{(\frac{\cosh r}{\tanh r})^{k_1+j_1-p-u}\ (-\sinh r)^{p+u}}{p!\
(k_1-p)!\ u!\ (j_1-u)!}\nonumber \\
&\times \frac{(n_2+p)!\ (m_2+u)!}{(n_1-k_1+p)!\ (m_1-j_1+u)!}\
\delta_{n_2-n_1,k_2-k_1}\ \delta_{m_2-m_1,j_2-j_1}\,,
\end{align}
which can also be expressed in terms of a product of values of
the
hypergeometric
function $_2F_1$, as given by Eq.~(\ref{Eq:PDC-Tensor}).



\begin{thebibliography}{99}
\bibitem{footnote1} We adopt $[\hat {x},\hat{ p}]=i$ for the position and
 momentum quadratures [$\hbar\equiv1$], whence $\alpha=(x+ip)/\sqrt 2$.
\end{thebibliography}
\end{document}